\documentstyle[epsf,12pt]{article}
\begin{document}
\begin{titlepage}
\begin{center}
\vspace*{3cm}

\begin{title}
\bold
{\Huge
Event-by-event cluster analysis \\

of  final states from  heavy ion collisions }
\end{title}

\vspace{2cm}

\begin{author}
\Large
K. FIA{\L}KOWSKI\footnote{e-mail address: uffialko@thrisc.if.uj.edu.pl},
R. WIT\footnote{e-mail address: wit@thrisc.if.uj.edu.pl}

\end{author}

\vspace{1cm}

{\sl M. Smoluchowski Institute of Physics\\ Jagellonian University \\

30-059 Krak{\'o}w, ul.Reymonta 4, Poland}

\vspace{3cm}

\begin{abstract}
We present an event-by-event analysis of  the cluster structure of final
multihadron states resulting from heavy ion collisions. A comparison of
experimental data with the states obtained from Monte Carlo
generators is shown. The analysis of the first available
experimental events suggests that the method is suitable for selecting
some different types  of events.
\end{abstract}

\end{center}



\vspace{1cm}

\noindent

May, 1999 \\

\end{titlepage}

\section{Introduction}
\par
 In many scenarios of heavy ion collisions, such as Quark Gluon Plasma
(QGP)[1] or Disoriented Chiral Condensate (DCC) [2] formation, one
expects an appearance of groups of particles, which differ somehow from
the majority of produced hadrons. In particular, such groups are often
 characterized by small momenta in the rest frame of the group forming a
sort of a cluster. Therefore it seems to be interesting to develop
procedures looking for clusters in single events.
\par We analyze the
cluster structure of events resulting from the heavy ion collisions and
from the Monte Carlo (MC) generators. In particular, we investigate, for
single events, the dependence of the multiplicity distributions of charged
hadrons in clusters on the parameter defining a cluster size. The
procedure is presented in next section. We employ the
cluster definition used in the procedure implementing the Bose-Einstein
(BE) effect in MC [3]. We consider only
clusters containing at least two hadrons and investigate the average
multiplicity and the second normalized factorial moment, defining them
with "two subtracted", i.e.  $\overline n -2$, and

\begin{equation}
\tilde{F}_2 = \frac {\overline {(n-2)(n-3)}}{(\overline n -2)^2}.
\end{equation}

\par
This  choice of moments is motivated by three facts.
First, in our procedure each cluster contains by definition at least one
hadron.  Second, the percentage of "one hadron clusters" is not expected
to obey any smooth multiplicity distribution (e.g. due to the
contributions of "leading particles"). Therefore only the
 distribution for $n \geq 2$ may be expected to follow from some
 smooth formula. Moreover, if we consider  the limit for $\tilde F_2$ at
 $\overline n -2 \rightarrow 0$  for Negative Binomial Distribution
 in $n-2$ with different values of the parameter $k$, we find it  is
 finite and depends on $k$.  Thus, in principle, our
 choice of moments allows to investigate the cluster structure in the
 limit of very small clusters, where  other definitions lead to
divergencies or to results independent of the shape of the distribution.
\par
In this note we work in a two-dimensional momentum space ($\eta -
\phi$) to be able to use the available experimental data of the KLM
collaboration [4].  In principle, however, the data with full measurement
of momenta are preferred, as they should allow to identify clusters better
suited for the analysis of QGP or DCC signals.
\par In the next section
 we present in more detail our procedure and the form in which the data
 are presented.  We discuss also some modifications and
 generalizations of our approach.  In the third section we compare the
results of our analysis for the real - and MC generated events. The last
section contains conclusions and outlook.

\section{Clustering procedure}
\par
To start the clustering procedure we have to define the distance in the
momentum space. Since in the data set [4] only the pseudorapidity $\eta$
 and the azimuthal angle $\phi$ are given, we use
\begin{equation}
d^2 = (\Delta \eta)^2 + (\Delta \phi)^2
\end{equation}
as a distance measure.
An introduction of different coefficients in front of
the two terms in this definition (with their ratio between 0.5 and 2.0)
does not change significantly the final results. Obviously, for data with
 full momenta measurement we would use some other measures, e.g. three
 momentum or four momentum difference squared.

\par In our procedure
originally each particle is considered as a single cluster. We fix the
value of a "cluster size parameter" $\epsilon $ and perform the clustering
procedure for all pairs of particles. The particle is assigned to a given
cluster if its distance $d^2$ to at least one particle from this cluster
is smaller than $\epsilon$. Obviously for very small  $\epsilon$ values
we have very few (or no) clusters with at least two particles. On the
other hand, for large $\epsilon$ values almost all particles must fall
into a single large cluster.
\par Our procedure is
different from the clustering procedure applied in Ref. [4]. We do not
define the "global cluster size" in the ($\eta -\phi$) space but consider
only distances between particles. Moreover, the results are analyzed
differently. In Ref. [4] one asks what is the percentage of particles in
clusters; we investigate the multiplicity distribution of particles in
different clusters for single events.  The results are significant if we
have a sufficiently large number of clusters (with at least two
particles). This limits the possible range of $\epsilon$ values, which
must be neither too small nor too large.  Thus our analysis is in fact
complementary to the "intermittency analysis" performed in Ref. [4], where
only very small bins in momentum space are relevant (corresponding to
small $\epsilon$ values).  At the lower end of our range most of the
clusters contain just two particles; at the higher one, as we shall see,
there are typically a few tens of particles in a cluster. To characterize
the muliplicity distribution we calculate for each event the average
multiplicity of a cluster and the second scaled factorial moment, as
defined in the Introduction.

\section{Results: comparison of MC with data}
\par
We performed the clustering procedure for the four KLM events presented in
 Ref. [4] and for the events obtained from the VENUS generator [5]. In
 addition we use the "random events" of similar multiplicities obtained
with a uniform random generation of $(\eta, \phi)$ points (SERENE events).
 The results are shown for the range of  $\epsilon$ values for which there
are at least 40 clusters (with more than one particle) in the event. We
checked that the results do not depend on the order in which the data are
 read in, as expected.
\par The average cluster multiplicity (or, to be more precise,
 its excess over the minimal value of 2) increases monotonically with
 $\epsilon$. It is interesting to note that the value of $\epsilon$ for
 which the increase becomes faster than linear is the same as a value for
 which the number of clusters with at least two particles starts to
 decrease (for smaller $\epsilon$ the decrease resulting from joining
 small clusters into the large ones is compensated by formation of small
 clusters from single particles).  As shown in Fig.1a, the increase is
faster for events with larger global multiplicity $N$.  Two lower data
sets correspond to $N = 742$ and $N=743$, and the two higher ones to
$N=926$ and $N=986$, respectively.  If the distance $\epsilon$ is scaled
by the average distance $\epsilon_0$ between neighbouring particles
(inversely proportional to N, $\epsilon_0 \simeq 12\pi/N$), differences
between events disappear as shown in Fig.1b.

\vspace{0.5cm}
\epsfxsize=15cm
\epsfbox{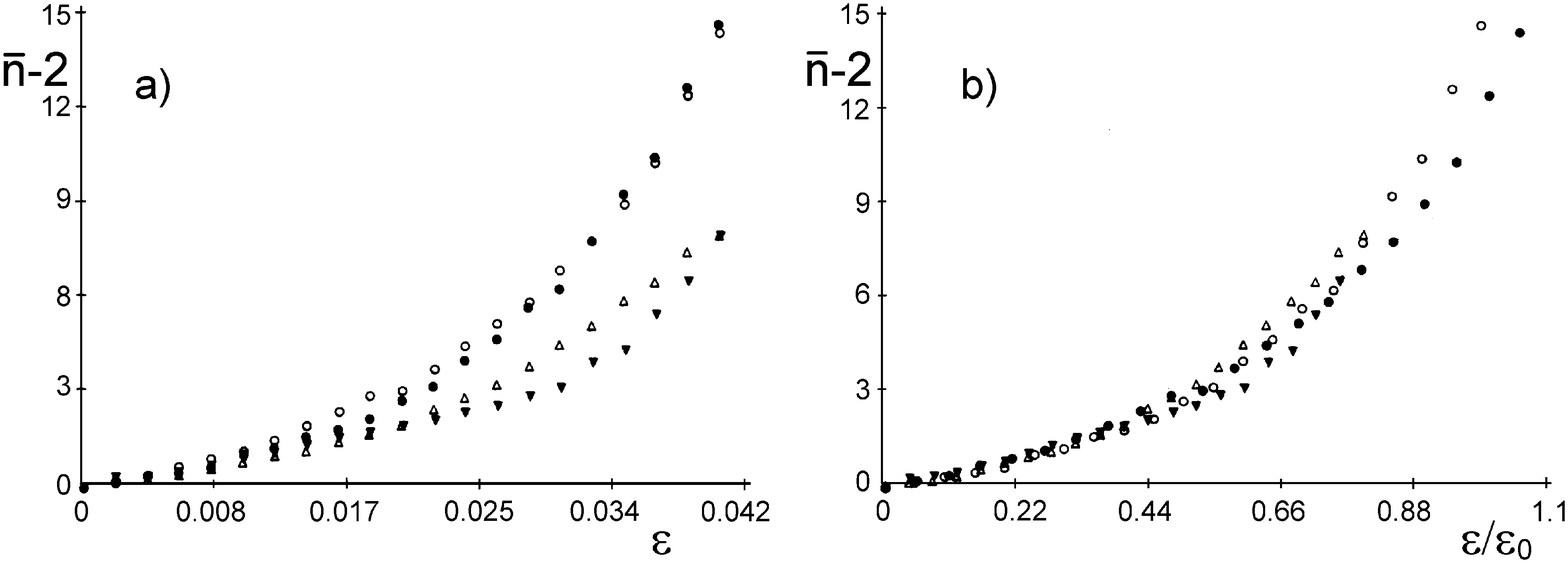}
\vspace{0.5cm}
\par {\bf Fig.1.} {\sl The
average hadron multiplicity of a cluster $\overline n -2$ (for clusters
with at least two particles) as a function of a)the parameter $\epsilon$
defining the maximal distance in the ($\eta - \phi$) space, for which two
particles are joined into a cluster, b)the same parameter scaled by
$\epsilon_0 = 12\pi/N$.  Data are taken from Ref. [4].  Four kinds of
symbols correspond to four different events. \\}
\par In the MC generated
events a similar increase occurs.  The corresponding plots (not presented
here) show no clear differences between the experimental and VENUS events,
although the latter ones have typically higher global multiplicities
resulting in higher average multiplicities of clusters. In random events
there seems to be a larger spread of data points for fixed global
multiplicity, but again no clear difference shows up.
\vspace{0.5cm}
\epsfxsize=10cm

~~~~~~~~~~~~~\epsfbox{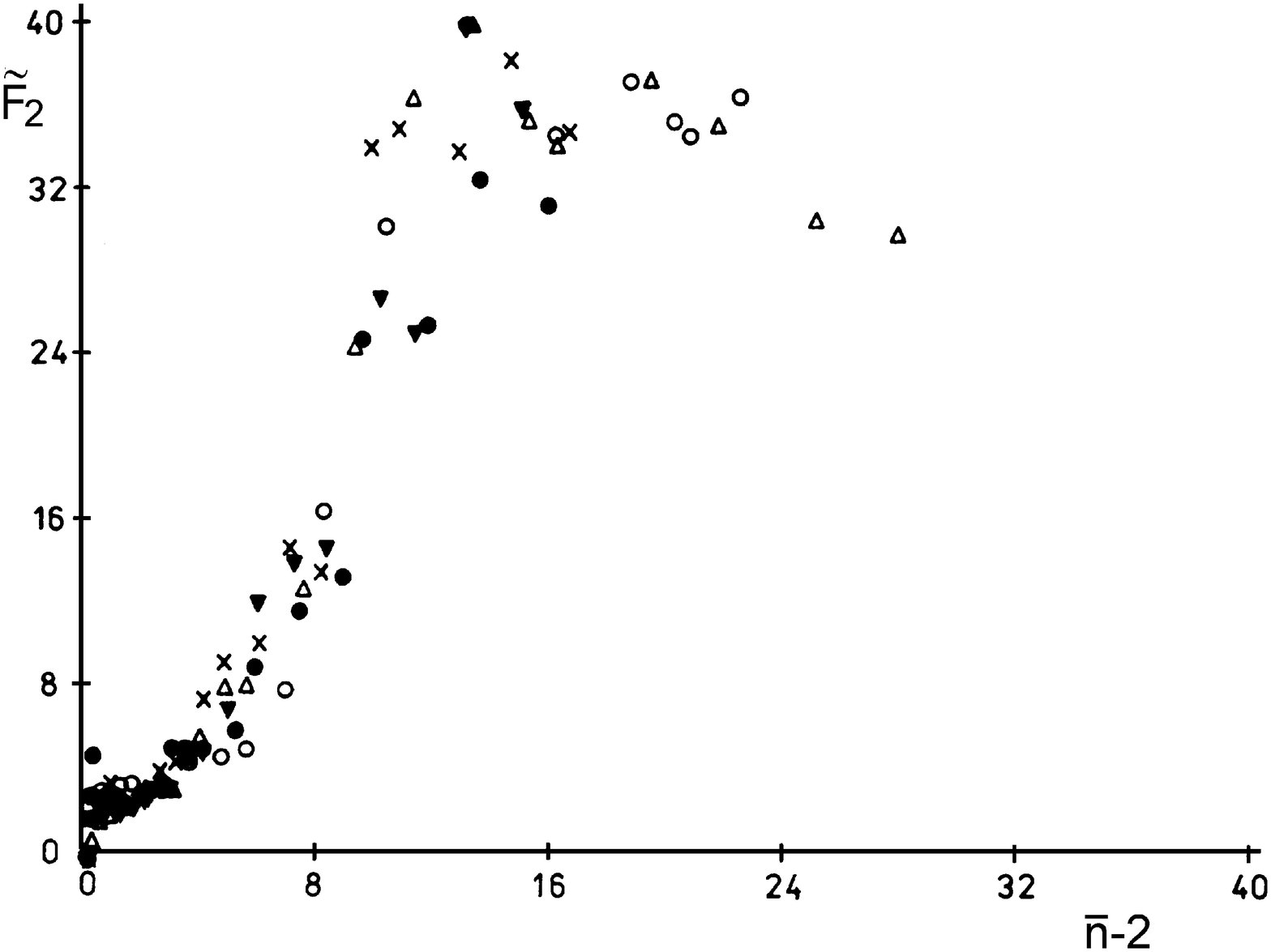}

\vspace{0.5cm}
\par
 {\bf Fig.2.} {\sl The second scaled factorial moment (Eq. (1)) as a
function of average multiplicity $\overline n -2$  for the events obtained
from the VENUS generator. Each symbol corresponds to one event.\\}

\par
On the other hand
the analysis of the second scaled factorial moment as a function of
average multiplicity shows a striking differentiation of events. For small
$\overline n -2$ (below 4) it rises irregularly to the values of a few
units. Then in all the VENUS events we see a much faster rise to the
values of about 30 - 40 as shown in Fig.2. For transparency only five out
of twenty analyzed events are shown.
\par
\vspace{0.5cm}
\epsfxsize=10cm

~~~~~~~~~~~~~\epsfbox{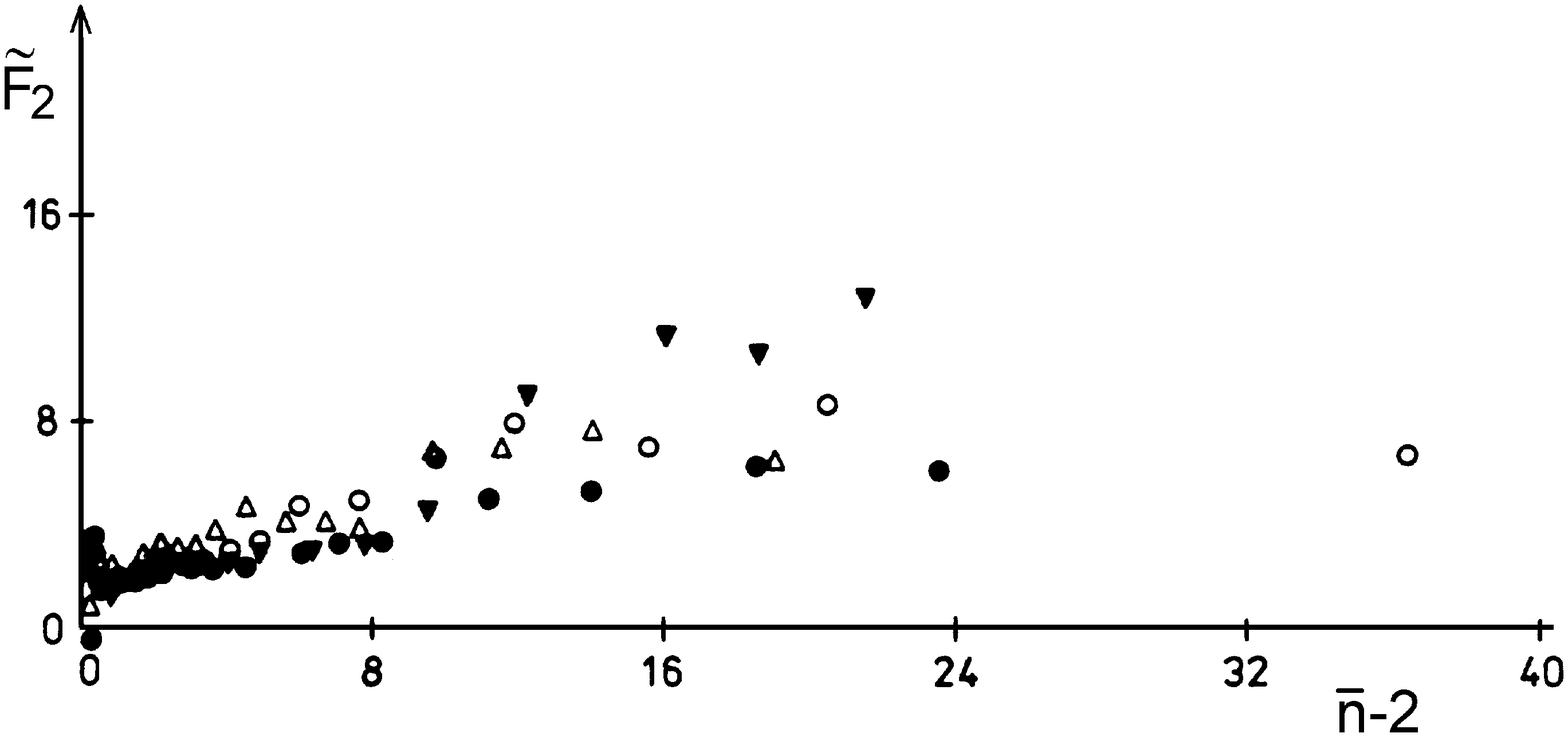}

\vspace{0.5cm}
\par
 {\bf Fig.3.} {\sl The second scaled factorial moment as a
function of average multiplicity $\overline n -2$  for the events obtained
from a uniform random numbers generator.  Each symbol corresponds to one
event.\\ }
\par
In the random events the rise
is much slower (Fig.3), although we selected multiplicities corresponding
to the maximal multiplicity of the VENUS events. Out of four KLM events
two seem to follow the "VENUS pattern" and two look like "SERENE
events", as seen in Fig.4.
\vspace{0.5cm}
\epsfxsize=10cm

~~~~~~~~~~~~~\epsfbox{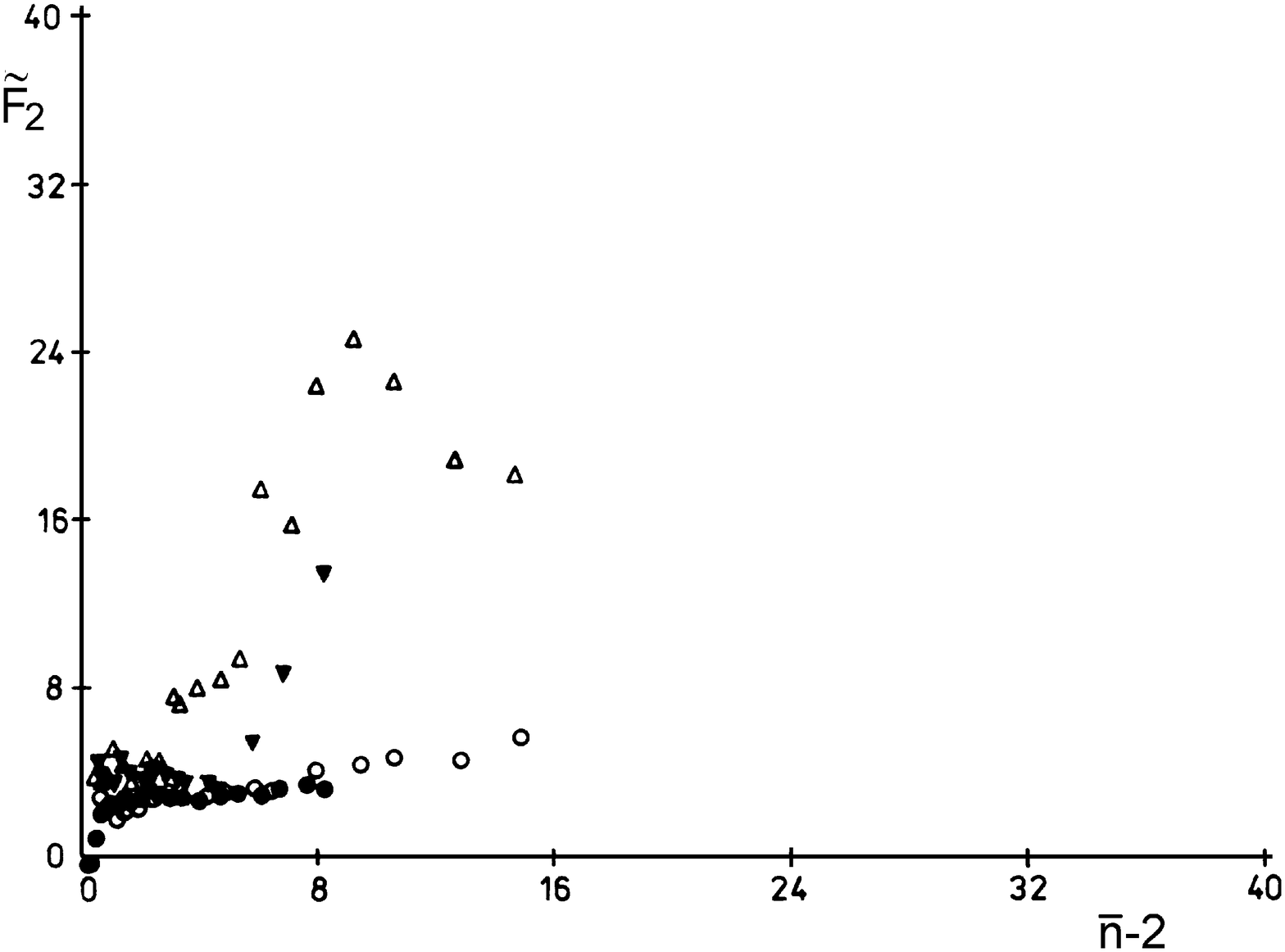}

\vspace{0.5cm}
\par
 {\bf Fig.4.} {\sl The second scaled factorial moment as a
function of average multiplicity $\overline n - 2$  for the KLM events.
Each symbol corresponds to one event.\\}
\par
The irregularity of $\tilde F_2$ for small
$\overline n -2$ shows that for the events with global multiplicity of the
order of $10^3$ we cannot draw any significant conclusions about the limit
$\overline n-2 \rightarrow 0$. Our choice of the definition
of $\tilde F_2$ "with 2 subtracted" (1) is found {\sl a posteriori} not
too relevant. However, the situation may be different at higher
multiplicities expected at RHIC energies.
\par
One should stress that large values of $\tilde{F}_2$ signal large
fluctuations of the cluster multiplicities (as compared to the average
value). The violent growth observed in VENUS events suggests an occurence
of one or a few clusters with multiplicity of the order of 100 when the
average fluctuates around 10. Apparently this does not occur in the
randomly sampled events and only two out of four KLM events show such
behaviour. It is interesting to note that these two events are the same
which show fast "intermittent" rise of the scaled factorial moments in
small bins of the phase-space; this rise was interpreted  as a
manifestation of "strong dynamical correlations" [4].

\par
It is, in fact, rather surprising that the same events seem to be
"special" in the limit of very small bins ("intermittency analysis") and
for a quite large average cluster size in our analysis.
\par
Even more surprising is the fact that in our analysis all the "VENUS
events" look like "special" KLM events. It may be so that fixing the
parameters of the VENUS generator by fitting the averages over many
experimental events induces such cluster structure in all single events,
whereas the data are more differentiated. It will be really interesting to
repeat this analysis on a much larger sample of events (both experimental
and MC generated) and for other choices of variables and clustering
procedures.
\section{Conclusions and outlook}
We have applied a simple cluster analysis to single events of heavy ion
collisions, both experimentally and MC generated. Analyzing the
multiplicity distributions of charged hadrons in clusters we found two
different patterns of behaviour already in the first four published
experimental events. On the contrary, all the VENUS generated events
follow one pattern. Thus the reasonable agreement of MC description with
the inclusive features of data (obtained by averaging over many events)
does not apply for single events.
\par We conclude that the cluster
analysis of single events may allow to select special classes of events
signaling, e.g., the formation of some exotic states (QGP, DCC). Obviously
our results present only a preliminary evidence; similar analysis should
be performed using different variables.  In particular, data with full
momentum measurements and particle identification should be compared with
the MC results. Different clustering procedures should also be introduced.
The results may help to improve our understanding of multiple production
in heavy ion collisions.

\section{Acknowledgements}
 B. Wosiek kindly provided the data published in Ref. [4]
and the "VENUS data" in a ready-to-analyze form. Technical assistance
by M.  Wit is acknowledged. We thank A. Bia\l{}as for some useful
remarks. The financial support of KBN grants \# 2 P03B 086 14 and \# 2
P03B 010 15 is gratefully acknowledged.  One of us (RW) is grateful for a
partial financial support by the KBN grant \# 2 P03B 044 12.

\section{References}
\noindent
[1] M. Jacob, J. Tran Thanh Van (eds.), {\sl Phys. Rep.} {\bf 88}, 321
(1982); E.V. Shuryak, {\sl Phys. Rep.} {\bf 115}, 151 (1984).\\
\noindent
[2] A. Anselm, M.G. Ryskin, {\sl Phys. Lett.} {\bf B266}, 482
(1991); J.D. Bjorken, {\sl Int. J. Mod. Phys.} {\bf A7}, 4189 (1992);
J.-P. Blaizot, A Krzywicki, {\sl Phys. Rev.} {\bf D46}, 426 (1992);
 K. Rajagopal, F. Wilczek, {\sl Nucl. Phys.} {\bf B399}, 395 (1993).\\
\noindent
[3] K. Fia{\l}kowski, R. Wit and J. Wosiek, {\sl Phys. Rev.} {\bf D58}
094013 (1998).\\
\noindent
[4] M.L. Cherry at al. (KLM Collaboration), {\sl  Acta Phys.
Polon.} {\bf 29}, 2129 (1998).\\
\noindent
[5] K. Werner, {\sl Phys. Rep.} {\bf 232}, 87 (1995).\\
\noindent
\end{document}